\documentclass{aa}
\usepackage{graphics}
\usepackage{amsmath}
\usepackage{amssymb}
\usepackage{natbib}
\newcommand{\comments}[1]{\bigskip\parbox[t]{0.9\linewidth}{\small{#1}}}

\newcommand{\e}{{\rm e}}
\newcommand{\half}{\tfrac{1}{2}}
\newcommand{\gccm}{\mathrm{\,g\,cm}^{-3}}
\newcommand{\sfs}{$^1\mathrm{S}_0$}
\newcommand{\sfp}{$^3\mathrm{P}_2$}
\newcommand{\UVU}{UV$_{14}$+UVII}
\newcommand{\UVT}{UV$_{14}$+TNI}
\newcommand{\HVpn}{HV$_{\rm pn}$}
\newcommand{\RHFpn}{RHF$_{\rm pn}$}
\newcommand{\Kpn}{G$^{\rm K240}_{\rm pn}$}

\newcommand{\KM}{G$^{\rm K240}_{\rm M78}$}
\newcommand{\KMu}{G$^{\rm K240}_{\rm M78u}$}

\begin{document}

\thesaurus{06(08.14.1; 08.05.3; 02.04.1; 02.13.1; 13.25.5)}

\title{Neutron superfluidity in strongly magnetic interiors of neutron stars
       and its effect on thermal evolution}
\author{Ch. Schaab\thanks{E-mail address: schaab@gsm.sue.physik.uni-muenchen.de}
  \and F. Weber \and M. K. Weigel} 
\institute{Institut f{\"u}r theoretische Physik,
  Ludwig-Maximilians Universit{\"a}t M{\"u}nchen, Theresienstr. 37,
  D-80333 M{\"u}nchen, Germany}

\date{Received: October 6, 1997 / Accepted March 31, 1998}

\titlerunning{Neutron superfluidity in strongly magnetic neutron stars}
\authorrunning{Schaab et al.}
\maketitle

\begin{abstract}
The possibility of a neutron $m=2$--superfluid in the interior of
neutron stars is investigated. This pairing state is energetically
favoured in strong magnetic fields ($H\sim 10^{16}-10^{17}$~G). Because of the
node in the angular-dependent energy gap along the field direction the
neutrino emissivity is only suppressed polynomially as function in
$T/T_{\rm c}$ instead of exponentially, as it is obtained for a
nodeless pairing state. The effect of this pairing state on the
thermal evolution of neutron stars is studied, and its outcome is
compared with the evolution of ``normal'', i.e. nodeless, superfluid
and non-superfluid neutron stars, and also with observations. We find
that particularly the predicted surface temperatures of the enhanced
cooling scenario considerably change and come into agreement with
temperatures deduced from observational data within the hydrogen
atmosphere model. Furthermore the surface temperature depends on the
magnetic field strength as an additional parameter aside from the
neutron star mass. The latter is however only operative in the case of
the intermediate cooling scenario.

\keywords{Stars: neutron -- Stars: evolution -- Dense matter --
  Magnetic fields -- X-rays: stars}
\end{abstract}

\section{Introduction} \label{sec:intro}

Comparison of soft X-ray and extreme UV observations with simulations
of the thermal evolution of neutron stars gives us a powerful tool for
investigating the interior of such stars, since young neutron stars
cool mainly via neutrino emission from there. The rate of neutrino
emission depends on a number of ingredients, among which the
composition and the superfluid pairing play the most important roles.

Differences in the composition of various models make it sensible to
distinguish between slow (or standard) and enhanced scenarios of
thermal evolution, depending on whether the dominant process is either
the modified or the direct Urca. As it was shown in \cite{Schaab95b}
\cite[see also][]{Voskresenskii86} the uncertainties in the degree of
modifications of the modified Urca rate by medium effects suggest a
third possible scenario, the intermediate cooling scenario.

It is generally believed that neutrons in the interior of a neutron
star form superfluid pairs in a \sfp -state below some critical
temperature $T_{\rm c}\sim 10^9$~K. Since this high density ($\rho
>\rho_0=2.8\times 10^{14}\gccm$) matter in $\beta$-equilibrium is not
directly accessible by laboratory experiments, one can only extrapolate
the two-body interaction to this regime. The pairing gaps however
depend very sensitively on the underlying microscopic model for the
effective nucleon--nucleon interaction. Its value and the density
range over which pairing occurs is therefore quite uncertain.  Another
uncertainty concerning the \sfp -pairing of neutrons refers to the
angular dependency of the gap. In isotropic matter the $m=0$--angular
dependency $\Delta^2(\vec k)\propto 1/3+\cos^2\theta$ is energetically
favoured \cite[]{Hoffberg70a}. Above some critical
magnetic field strength $H_{\rm c}\sim 10^{16}-10^{17}$~G however the free
energy is minimised by a $m=2$--dependency $\Delta^2(\vec
k)\propto \sin^2\theta$, where $\theta$ is the angle between the
quasiparticle momentum $\vec k$ and the magnetic field direction
\cite[]{Muzikar80}.  This different angular dependency
turns out to be important for the cooling behaviour of superfluid
neutron stars, since the $m=2$--angular dependency has nodes at
$\theta=0,\pi$. The available phase space of
the various processes keeps therefore undiminished in the direction of
the magnetic axis, and the reaction rates are only suppressed
polynomially instead of exponentially
\cite[]{Anderson61,Levenfish94a}.

The observed surface magnetic fields of neutron stars ranges from
$B=1.7\times 10^8$~G (PSR B1957+20) up to $2.1\times 10^{13}$~G (PSR
B0154+61), with a median value of about $10^{12}$~G. These values are
deduced from the observed spin-down velocity of 558 pulsars by
assuming the magnetic dipole braking model \cite[see the pulsar catalog by][]{Taylor93a}). Because of the highly conducting
core the surface may hide an even stronger magnetic field in the
interior \cite[]{Muslimov96a}. By applying the Newtonian scalar virial
theorem \cite{Lai91a} showed that the interior field
may be as high as $10^{18}$~G. Works on the origin and evolution of
the interior magnetic field show that at least field strengths of
$H=10^{15}$ to $10^{17}$~G are possible
\cite[]{Bisnovatyi92a,Bisnovatyi93a,Duncan92a,Thompson93a}, which is in
the range of the critical field strength for $m=2$--superfluidity.

Besides the neutrons also the protons supposedly pair in the interior
of a neutron star. Since they probably form a type II-superconductor,
the magnetic field is confined to flux tubes, whose radial dimension
is given by the London penetration depth, $\lambda\sim
40\text{--}100$~fm. The magnetic field inside the flux tubes is
therefore of the order of $H\approx\Phi_0\ln(\lambda/
\xi)/2\pi\lambda^2 \sim 10^{14}\text{--}10^{16}$~G, where
$\Phi_0\approx 2.1\times 10^{-7}$ G~cm$^2$ is the flux quantum and
$\xi\sim 10$~fm is the coherence length. Even if the bulk of the
neutrons do not experience a field above the critical field $H_{\rm
c}$, the field inside the flux tubes may be large enough. This would
be sufficient for a considerable modification of the thermal
properties of a neutron star.

In this work we will study the effect of nodes of the neutron gap
energy along the magnetic axis on the thermal evolution of neutron
stars and compare the results with neutron star models with \sfp
-pairing in the $m=0$--state and also with observed data. We will
start with the underlying physics by discussing the energetically
favoured pairing state in \sfp -superfluidity and the suppression of
the most important neutrino emission processes
(Sect. \ref{sec:pairing} and
\ref{sec:neutrino}).  In Sect. \ref{sec:models} we classify our models
into 16 scenarios. The outcome of the simulations for these scenarios
is compared with observations in Sect. \ref{sec:res}. Section
\ref{sec:disc} contains the conclusions and the discussion.

\section{\sfp -pairing} \label{sec:pairing}

Neutrons are believed to form \sfs -superfluid pairs in the density
range between neutron drip and the saturation density of normal
nuclear matter. At supernuclear densities, i.e. in the core regime of
neutron stars, the repulsive interaction becomes dominant and the \sfs
-superfluidity gap closes. However, the tensor and spin-orbit
interaction become attractive at such densities and the formation of
\sfp -neutron pairs becomes possible. The ordering parameter for \sfp
-pairing is a complex, traceless, symmetric 3x3-matrix $\bf A$. It can
be determined in vicinity of the transition temperature $T_{\rm c}$ by
minimising the Ginzburg-Landau free-energy functional $\Omega[{\bf
A}]$. This was done by \cite{Muzikar80}, who found
that for a magnetic field strength
\begin{equation} \label{eq:crit.field}
  H > H_{\rm c}(T) = 3\times 10^{17} \frac{k_{\rm B}T_{\rm c}}{1{\rm ~MeV}}
    \left( 1-\frac{T}{T_{\rm c}}\right) {\rm ~G}
\end{equation}
the energy gap for quasiparticle of momentum $\vec p$ is
\begin{equation} \label{eq:gap1}
  |\Delta_{m=2}(\hat p)|^2 = \frac{3}{8\pi}\Delta_{m=2}^2(T)\sin^2\theta~.
\end{equation}
Here, $k_{\rm B}$ denotes the Boltzmann-constant, $\theta$ the angle
between the magnetic field $\vec H$ and quasiparticle momentum $\vec
p$, $\hat p$ the unit vector in direction of $\vec p$, and
$\Delta_{m=2}^2(T)$ a temperature dependent factor. For vanishing magnetic
fields the free energy functional is minimised by the gap function
\begin{equation} \label{eq:gap2}
  |\Delta_{m=0}(\hat p)|^2 = \frac{1}{16\pi}\Delta_{m=0}^2(T)(1+3\cos^2\theta')~,
\end{equation}
where $\theta'$ is the angle between $\vec p$ and some quantisation
axis. As we have already noted the gap function \eqref{eq:gap1}
vanishes along the magnetic axis at $\theta=0,\pi$, whereas the gap
function \eqref{eq:gap2} is nodeless.  For magnetic fields $0<H<H_{\rm
c}(T)$ the minimising gap function is a superposition of Eqs.
\ref{eq:gap1} and \ref{eq:gap2}, which is still nodeless.

The critical field strength vanishes at the critical temperature
$T_{\rm c}$ and increases to
\begin{equation}
  H_{\rm c}(T=0) = 3\times 10^{17} \frac{k_{\rm B}T_{\rm c}}{1\,{\rm
      MeV}} \,{\rm G}
\end{equation}
at $T=0$. In the pairing model of \cite{Hoffberg70a} the maximum energy gap is, for instance,
$\Delta_{\rm max}\approx 4.5$~MeV. This yield a critical field
strength $H_{\rm c}(T=0)\approx 2\times 10^{17}$~G. If the inner
magnetic field is smaller than $H_{\rm c}(T=0)$, first a
$m=2$--superfluid is formed at $T_{\rm c}$ and then at
\begin{equation} \label{eq:crit.temp2}
 T_{\rm c,2} = T_{\rm c}\left( 1-\frac{H}{H_{\rm c}(T=0)}\right)
\end{equation}
a nodeless $m=0$--superfluid is formed.

\section{Neutrino emissivity in superfluid matter}
\label{sec:neutrino}

In the vicinity of the Fermi surface ($|p-p_{\rm F}|\ll p_{\rm F}$)
the one-particle energy can be approximated by \cite[see, e.g.,][]{Lifshitz80a}
\begin{equation}
  E({\vec p}) = \mu \begin{cases}
   -\sqrt{|\Delta(\hat p)|^2+v_{\rm F}^2(p-p_{\rm F})^2} & \text{for $p<p_{\rm F}$,} \\
   +\sqrt{|\Delta(\hat p)|^2+v_{\rm F}^2(p-p_{\rm F})^2} & \text{for $p\ge p_{\rm F}$,}
  \end{cases}
\end{equation}
where $v_{\rm F}$, $p_{\rm F}$, and $\mu$ is the Fermi velocity, 
momentum, and energy, respectively. The superfluidity creates a
gap of width $2|\Delta(\hat p)|$ in the energy spectrum around the
Fermi momentum $p_{\rm F}$. Excitations of particles above the Fermi
energy are therefore suppressed for $k_{\rm B}T<|\Delta(\hat p)|$. We
shall study the two possible cases of triplet pairing of neutrons
which occur (or might occur) in neutron stars' interior (see Sect.
\ref{sec:pairing}). The gap function $|\Delta_{m=0}(\hat p)|$ in the case
of nodeless pairing (Eq. \ref{eq:gap2}) has the approximate form
\cite[]{Levenfish94a}
\begin{multline}
  |\Delta(\hat p)_{m=0}| = \sqrt{1-\frac{T}{T_{\rm c}}}\left(
  0.7893+1.188\frac{T_{\rm c}}{T}\right)k_{\rm B}T \\
  \times \sqrt{1+3\cos^2\theta},
\end{multline}
whereas gap \eqref{eq:gap1} gives
\begin{multline}
  |\Delta(\hat p)_{m=2}| = \sqrt{1-\left(\frac{T}{T_{\rm c}}\right)^4} \\
  \times \left( 2.030-0.4903\left(\frac{T}{T_{\rm c}}\right)^4 
  +0.1727\left(\frac{T}{T_{\rm c}}\right)^8\right)
  \frac{T_{\rm c}}{T}k_{\rm B}T\sin\theta.
\end{multline}

\begin{figure}
\resizebox{\hsize}{!}{\includegraphics{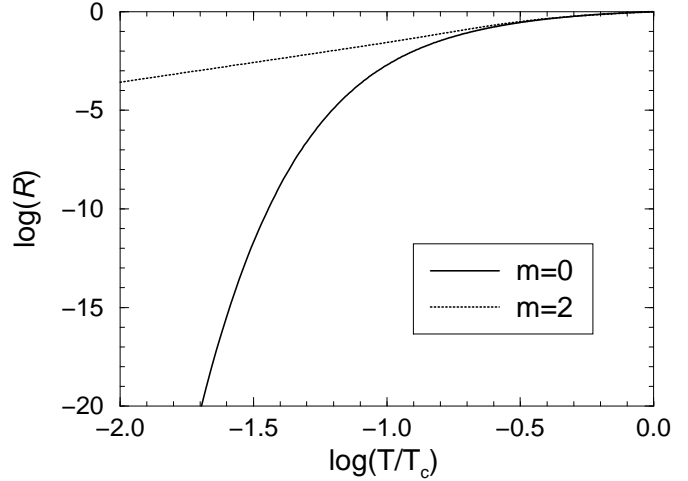}}
\caption[]{
  Suppression factor ${\mathcal R}_{\rm n}^{\rm d}$ 
  of the direct Urca process for the
  two cases of triplet pairing. \label{fig:supr}}
\end{figure}

The suppression of the neutrino emissivity $\epsilon_\nu$ by
superfluid neutrons can be expressed by a factor
\begin{equation}
  {\mathcal R_{\rm n}} = 
  \frac{\epsilon_\nu^{\rm sf}}{\epsilon_\nu^{\rm n.sf}},
\end{equation}
which was calculated for a couple of processes in \cite{Levenfish94a} and \cite{Yakovlev95a}. Let us consider for example the direct Urca
process \cite[]{Boguta81a,Lattimer91}
\begin{equation} \label{eq:dirUrca}
  {\rm n} \rightarrow {\rm p} + {\rm e}^- + \bar\nu_{\rm e}.
\end{equation}
The suppression factor can analytically be calculated only in the
limiting case $T\rightarrow 0$ \cite[]{Levenfish94a}
\begin{equation} \label{eq:supr1}
  {\mathcal R}_{\rm n}^{\rm d} = \begin{cases}
    \frac{126}{457\pi^5}\sqrt{\frac{1}{3}} v^5 \e^{-v} 
    & \text{for case $m=0$,} \\
    \frac{6029\pi^2}{5484} v^{-2} 
    & \text{for case $m=2$,} \end{cases}
\end{equation}
where the dimensionless parameter $v$ is given by
\begin{equation}
  v = \begin{cases}
  \sqrt{\frac{1}{16\pi}}\frac{\Delta(T)}{k_{\rm B}T} & \text{for case $m=0$,} \\
  \sqrt{\frac{3}{8\pi}}\frac{\Delta(T)}{k_{\rm B}T} & \text{for case
    $m=2$.}
  \end{cases}
\end{equation}
By numerical fitting \cite{Levenfish94a} obtain
\begin{equation}
  {\mathcal R}_{\rm n}^{\rm d} = \begin{cases}
  \left(0.2546+\sqrt{(0.7454)^2+(0.1284v)^2}\right)^{5} & \\
  \quad\times\exp\left(2.701-\sqrt{(2.701)^2+v^2}\right) 
  & \text{for $m=0$,} \\
  \frac{\frac{1}{2}+(0.09226v)^2}{1+(0.1821v)^2+(0.16736v)^4} & \\
  \quad +\frac{1}{2}\exp\left(1-\sqrt{1+(0.4129v)^2}\right) & \text{for 
    $m=2$.}
  \end{cases}
\end{equation}
As can be deduced from Eq.  \ref{eq:supr1}, the suppression factors of
the nodeless case $m=0$ decreases with decreasing temperature much
more rapidly than the suppression factor for case $m=2$ (see Fig.
\ref{fig:supr}). This is caused by the two nodes of the gap function
at the intersections of the Fermi surface with the magnetic axis.

Besides the direct Urca process we considered a couple of other
processes (see Table \ref{tab:neutrino}). Following \cite{Levenfish96a} we approximate the suppression factors
for these processes by 
\begin{equation}
  {\mathcal R}^1_{\rm n}(v) = {\mathcal R}^{\rm d}_{\rm n}(v),
\end{equation}
if only one neutron is involved (proton branch of modified Urca
process, electron-neutron, and neutron-proton scattering), and by
\begin{equation}
  {\mathcal R}^2_{\rm n}(v) = {\mathcal R}^{\rm d}_{\rm n}(2v),
\end{equation}
if two neutrons are involved (neutron branch of modified Urca process
and neutron-neutron scattering).

\begin{table*}[tbp]
\caption[]{Neutrino processes considered in the cooling
    simulations \label{tab:neutrino}}
\begin{tabular}{ll}
\hline
Process & Emissivity from Ref. \\
\hline
pair-, photon-, plasma-processes & \cite{Itoh89} \\
bremsstrahlung in the crust      & \cite{Yakovlev96a,Haensel96a} \\
bremsstrahlung in the core: & \\
\quad n-n scattering 		& \cite{Friman79} \\
\quad n-p scattering 		& \cite{Friman79} \\
\quad p-p scattering 		& \cite{Friman79} \\
\quad e$^-$-n scattering	& \cite{Kaminker97a} \\
\quad e$^-$-p scattering	& \cite{Kaminker97a} \\
\quad e$^-$-F scattering	& \cite{Kaminker97a} \\
superfluid pair breaking and formation
				& \cite{Voskresenskii87a,Schaab95b}; \dag \\
modified nucleon Urca: & \\
\quad neutron branch		& \cite{Friman79} \\
\qquad or alternatively		& \cite{Voskresenskii86,Schaab95b} \\
\quad proton branch		& \cite{Yakovlev95a} \\
direct nucleon Urca   		& \cite{Lattimer91} \\
direct hyperon Urca   		& \cite{Prakash92} \\
\hline
\end{tabular}
\comments{\dag : In the case of \sfp -pairing the emissivity was
recalculated by averaging over the angle $\theta$ between
quasiparticle momentum $\bf p$ and quantisation axis
(s. Eqs. \ref{eq:gap1}, \ref{eq:gap2})}
\end{table*}
 
The suppression factors for proton superconductivity are given in
\cite{Yakovlev95a}. If both neutrons and proton are superfluid we
approximate the combined suppression factor ${\mathcal R}_{\rm np}$ by
the minimum of the factors ${\mathcal R}_{\rm n}$ and ${\mathcal
R}_{\rm p}$ obtained by neglecting the superfluidity of the other
nuclear species \cite[cf.][]{Levenfish96a}:
\begin{equation} \label{eq:combined}
  {\mathcal R}_{\rm np} \approx 
  \min({\mathcal R}_{\rm n},{\mathcal R}_{\rm p}).
\end{equation}
We account also for the suppression of the heat capacity
\cite[]{Levenfish94b} and of the heat conductivity
\cite[]{Gnedin95a}. Thought the factors are similar as in the case of
neutrino processes, the effect on the thermal history is only small
compared to the effect of the suppression of the neutrino
emissivity. In the case of the heat capacity at least electrons and
possibly also non superfluid protons and hyperons still contribute to
the total heat capacity, whereas the heat conductivity is maintained
by electron-electron, electron-muon, and muon-muon scattering
processes.

\section{Cooling scenarios} \label{sec:models}

The general relativistic equations of stellar structure and thermal
evolution \cite[cf.][]{Thorne77} were numerically solved via an
implicit finite difference scheme by a Newton-Raphson algorithm (see \citeauthor{Schaab95a}, \citeyear{Schaab95a} for more details). The outcome of the
simulations depend on a couple of ingredients which are partly rather
uncertain. If one took all these uncertainties into account, one would
get a huge number of models, which are difficult to survey. However
all these models can be divided into four distinct classes depending
on the dominant neutrino process \cite[]{Schaab95b}. These are:
\begin{description}
\item[Class A -- Standard cooling:] This class consists of slowly cooling models
  which do not allow for any enhanced cooling process (like the direct
  Urca process).
\item[Class B -- Intermediate cooling:] The emission rate of the
  modified Urca process is increased by medium effects
  \cite[]{Voskresenskii86,Schaab95b}. This leads to a
  considerably faster cooling during the neutrino cooling era.
\item[Class C -- Enhanced nucleon cooling:] The third class allows for the direct
  nucleon Urca process (maybe in combination with pion or kaon
  condensation) yielding an even faster cooling than in the
  intermediate cooling scenario.
\item[Class D -- Enhanced hyperon cooling:] The last class allows
  additionally or exclusively for the direct hyperon Urca process,
  which is not suppressed by nucleon superfluidity in contrast to the
  nucleon processes\footnote{The ability of hyperons to form
  superfluid or superconducting pairs has only recently been
  investigated \cite[]{Balberg97b}. We shall assume here that hyperons
  cannot pair and address this problem in a forthcoming work
  \cite[]{Schaab98b}.}.
\end{description}
The possibility of the direct nucleon and hyperon Urca process depends
on the composition in the interior of the star and hence on the
underlying equation of state and on the central density of the star.
The equations of state can therefore also be divided into these four
classes, if the mass of the neutron star model is fixed to some value
(we selected the canonical value $M=1.4M_\odot$). Such a division is done in Table
\ref{tab:eos}. We used a broad collection of equations of state which
comprises relativistic field-theoretical models, one non-relativistic
Thomas-Fermi model, as well as non-relativistic Schr{\"o}dinger-based
ones. The first two classes, the standard and the intermediate cooling
classes, base on the same set of equations of state, but on different
neutrino emission rates of the modified Urca process. The division in
two different classes seems sensible, since the rate of this process
depends quite sensitively on the choice of the microscopic parameters
-- as, for instance, the effective pion gap -- which are unfortunately
rather uncertain.

\begin{table*}
\caption[]{Classification of equations of state into the four cooling
  classes defined in the text. The canonical neutron star
  mass $M=1.4M_\odot$ is assumed. The classification into the
  standard and intermediate cooling class depends on whether or not
  medium modification of the modified Urca process are taken into
  account. \label{tab:eos}}
\begin{tabular}{cccccc}
\hline
Class & EOS & Baryonic & Interaction & Many-body & Ref. \\
      &     & composition &          & approach &      \\
\hline
A/B & \UVU    & p,n & 2 nuclei potential Urbana V$_{14}$ and
    & NRV     & 1 \\
    &         &     & 3 nuclei potential Urbana VII  & & \\
A/B & \UVT    & p,n & 2 nuclei potential Urbana V$_{14}$ and
    & NRV     & 1 \\
    &         &     & density dependent terms TNI    & & \\
A/B & TF      & p,n & 2 nuclei potential TF
    & NRTF    & 2 \\
\hline
C   & \HVpn   & p,n & exchange of $\sigma,\omega,\rho$ mesons
    & RH      & 3 \\
C   & \RHFpn  & p,n & exchange of $\sigma,\omega,\rho,\pi$ mesons
    & RHF     & 4 \\
C   & \Kpn    & p,n & exchange of $\sigma,\omega,\rho$ mesons
    & RH      & 5 \\
\hline
D   & HV      & p,n,$\Lambda$,$\Sigma^{\pm,0}$,$\Xi^{0,-}$
    & exchange of $\sigma,\omega,\rho$ mesons
    & RH      & 3 \\
D   & RHF1    & p,n,$\Lambda$,$\Sigma^{\pm,0}$,$\Xi^{0,-}$,$\Delta$
    & exchange of $\sigma,\omega,\rho,\pi$ mesons
    & RHF     & 4 \\
D   & RHF8    & p,n,$\Lambda$,$\Sigma^{\pm,0}$,$\Xi^{0,-}$,$\Delta$
    & exchange of $\sigma,\omega,\rho,\pi$ mesons
    & RHF     & 4 \\
D   & \KM     & p,n,$\Lambda$,$\Sigma^{\pm,0}$,$\Xi^{0,-}$
    & exchange of $\sigma,\omega,\rho$ mesons
    & RH      & 5 \\
D   & \KMu    & p,n,$\Lambda$,$\Sigma^{\pm,0}$,$\Xi^{0,-}$
    & exchange of $\sigma,\omega,\rho$ mesons
    & RH      & 5 \\
\hline
\end{tabular}
\comments{Abbreviations: NRV: non-relativistic variational method, 
  NRTF: non-relativistic Thomas-Fermi model, RH: relativistic Hartree
  approximation, RHF: relativistic Hartree-Fock
  approximation. References: 1: \cite{Wiringa88}, 2:
  \cite{Strobel96a}, 3: \cite{Weber89}, 4: \cite{Huber97a}, 5:
  Glendenning, priv. comm.}
\end{table*}

As already mentioned in Sect. \ref{sec:intro} the second important
uncertainty concerns the superfluidity of neutrons in the interior of
the star. Again we classify the possible models into four classes:
\begin{description}
\item[Class 1:] Neutrons pair in the nodeless \sfp -state ($m=0$).
\item[Class 2:] The inner, uniform magnetic field strength $H$ 
  exceeds its critical value $H_{\rm c}(T=0)\approx 2\times 10^{17}$~G
  (see Eq. \ref{eq:crit.field}) even for zero temperature. The
  neutrons therefore form a $m=2$--superfluid.
\item[Class 3:] The inner magnetic field is set to $H=
  10^{17}{\rm\, G}<H_{\rm c}$. The critical field strength is
  therefore exceeded for temperatures $T_{\rm c,2}\leq T\leq T_{\rm
  c}$ only.
\item[Class 4:] Either the critical temperature of neutron pairing is
  very small \cite[e.g.][]{Elgaroy96c}, or the superfluid phase does
  not extend over the whole interior of the neutron star
  \cite[e.g.][]{Takatsuka72}.
\end{description}
For the classes 1--3 we used the gap energy by
\cite{Hoffberg70a}\footnote{\cite{Hoffberg70a} estimated the gap
energy of \sfp --pairing by using the phase shift data of
neutron--neutron scattering and assuming the effective nucleon mass to
be equal to its bare mass. The obtained gap energy serves therefore as
an upper limit and is used here to demonstrate the effect of
$m=2$--pairing.}. For class 4 we used the models by \cite{Elgaroy96c}
and \cite{Takatsuka72}. The angle-averaged energy gaps of \sfp
-pairing in the $m=0$-- and $m=2$--states are almost equal
\cite[]{Elgaroy96b}. We can therfore approximate the energy gap for
the $m=2$--state, which is not given by all authors, by
$\Delta_{m=2}(T)\approx\sqrt{\half}\Delta_{m=0}$ \cite[]{Amundsen85}.

The cooling simulations are performed by running the numerical code
for all combinations of parameters which are in agreement with the
respective cooling scenario (a detailed list of all parameters can be
found on the Web:
http://www.physik.uni-muenchen.de/sektion/suessmann /astro/cool/schaab.0797/input.html). This
procedure leads to cooling bands rather than individual cooling
tracks, by means of which we demonstrate the uncertainty inherent in
the input parameters. By combining the four classes of equations of
state A--D with the four classes of superfluidity 1--4 we get a
collection of sixteen scenarios whose behaviour will be studied in the
next section.

We shall study the case where the magnetic field exceeds $H_{\rm
c}(T=0)$ only inside the flux tubes of the proton type II-superconductor
separately. In this case (labeled as class 2$\gamma$), the suppression
factor $\mathcal R$ can be expressed by the combination
\begin{equation} \label{eq:gamma.suppr}
  {\mathcal R} = \gamma{\mathcal R}_{\rm in}
  +(1-\gamma){\mathcal R}_{\rm out}
\end{equation}
of the suppression factors ${\mathcal R}_{\rm in}$, ${\mathcal R}_{\rm
out}$ inside and outside the tubes. $\gamma$ is the volume fraction
occupied by the proton flux tubes. Inside the tubes we assume that the
magnetic field exceeds the critical field strength $H_{\rm
c}(T=0)$. Because protons are unpaired inside, the suppression factor
${\mathcal R}_{\rm in}$ is equal to the suppression factor for 
neutron $m=2$--pairing. Since the magnetic field is constrained to the flux
tubes, neutron pair in a nodeless $m=0$-state outside the
tubes. ${\mathcal R}_{\rm out}$ accounts for both superfluids, the
nodeless neutron superfluid and the singlet proton superfluid by the
approximate expression of Eq. (\ref{eq:combined}). The volume fraction
$\gamma$ is determined by the relation
\begin{equation} \label{eq:gamma.factor}
  \gamma = \frac{H_{\rm ex}}{H_{\rm tube}}
\end{equation}
of the bulk field strength $H_{\rm ex}$ and the field strength
inside the proton flux tubes $H_{\rm tube}$, which we set equal to the
critical field strength $H_{\rm c}(T=0)\approx 2\times 10^{17}$~G.

\section{Results and comparison with observations} \label{sec:res}

\begin{figure}
\resizebox{\hsize}{!}{\includegraphics{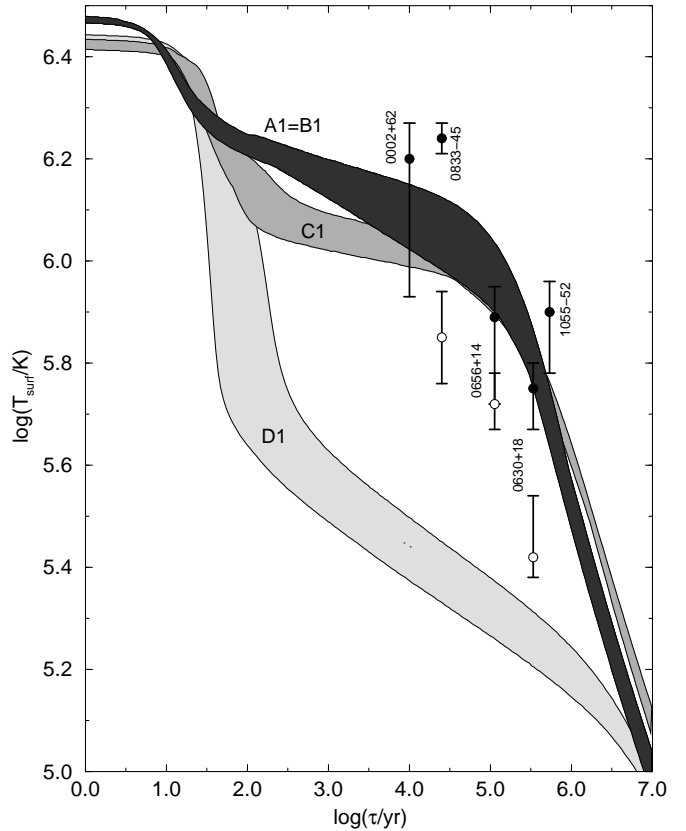}}
\caption[]{
  Thermal evolution of a $M=1.4M_\odot$ neutron star with
  superfluidity class 1. The bands correspond to the four classes of
  the equation of states as described in
  Sect. \ref{sec:models}. \label{fig:cool1}}
\end{figure}

\begin{figure}
\resizebox{\hsize}{!}{\includegraphics{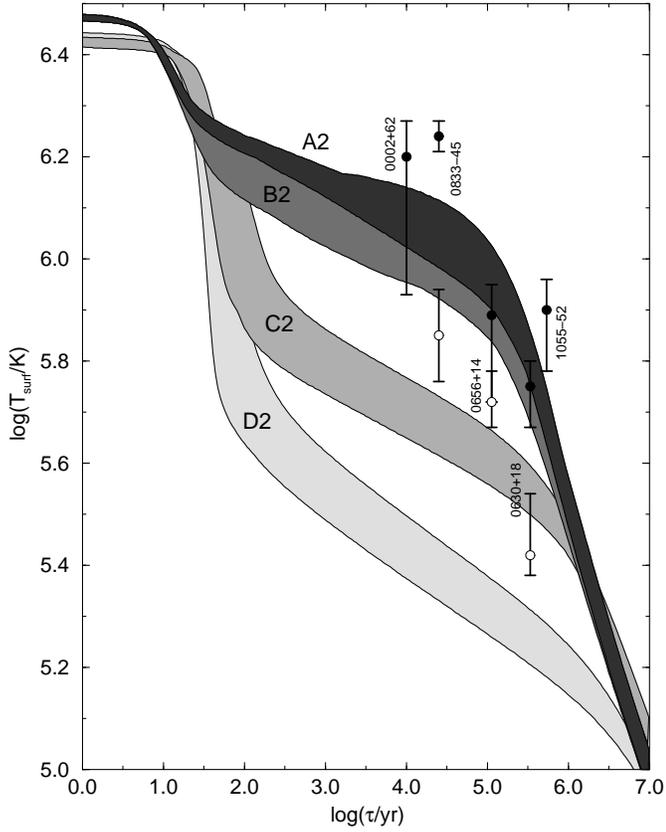}}
\caption[]{
  Same as Fig. \ref{fig:cool1} but for superfluidity class 2. \label{fig:cool2}}
\end{figure}

\begin{figure}
\resizebox{\hsize}{!}{\includegraphics{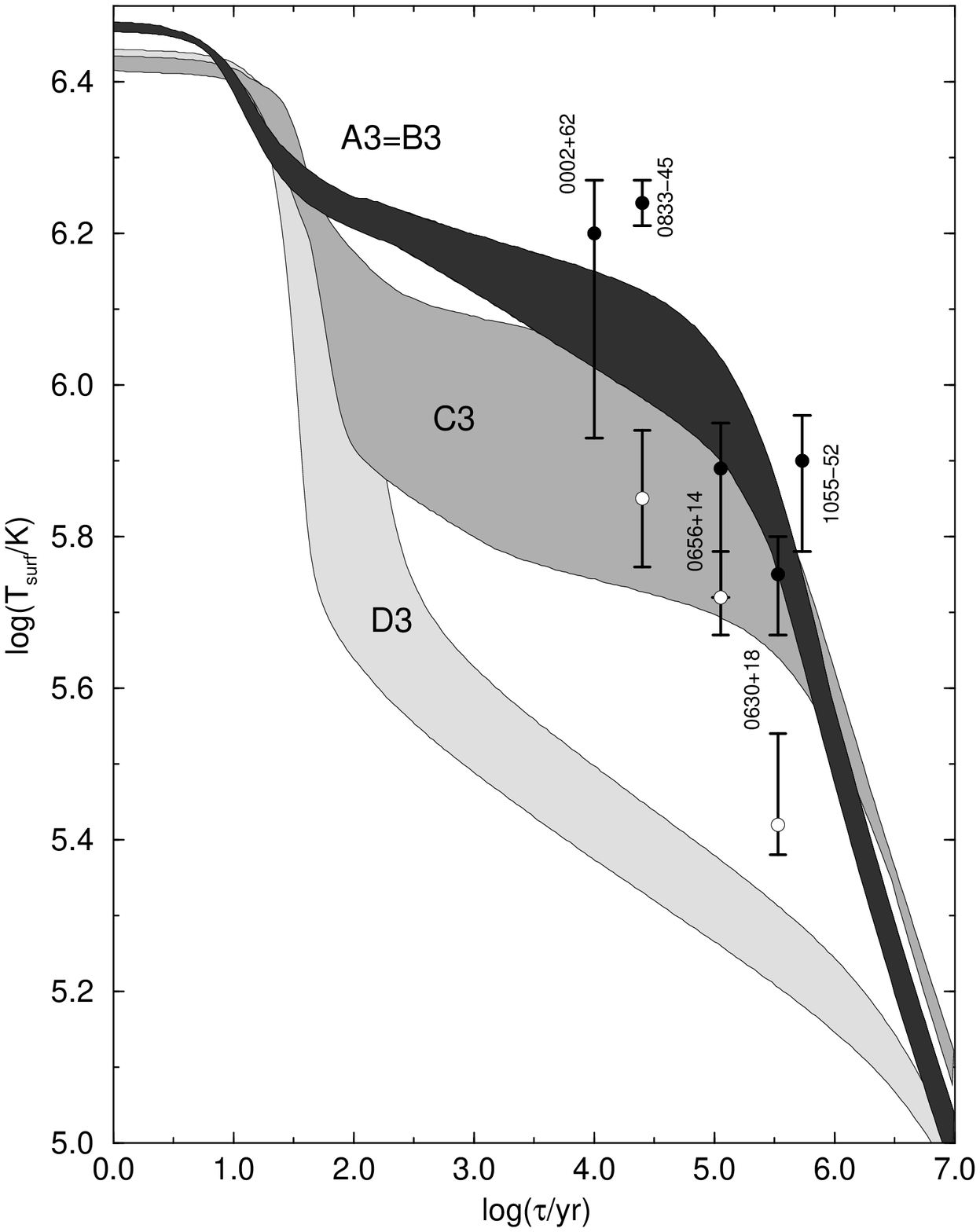}}
\caption[]{
  Same as Fig. \ref{fig:cool1} but for superfluidity class 3. \label{fig:cool3}}
\end{figure}

\begin{figure}
\resizebox{\hsize}{!}{\includegraphics{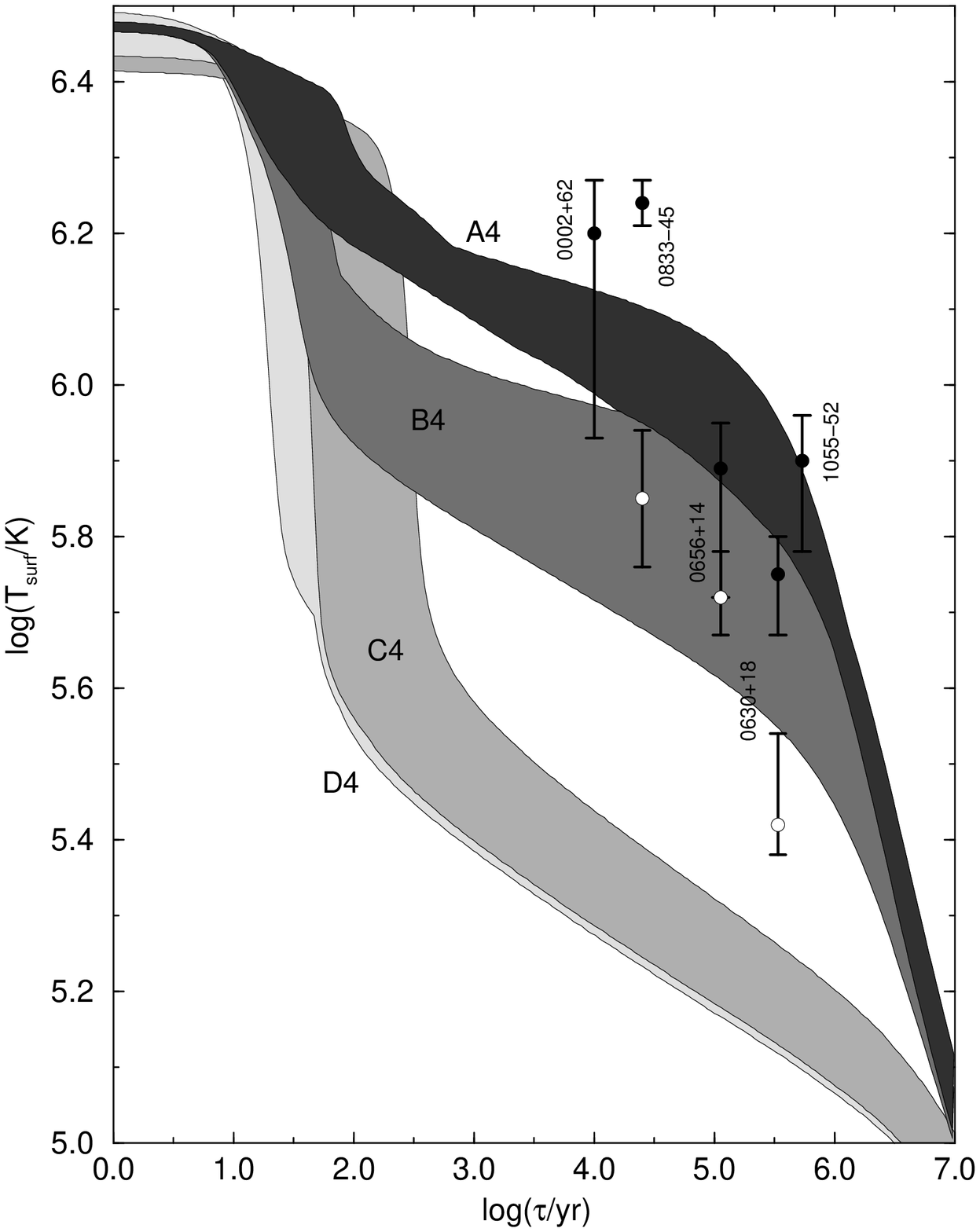}}
\caption[]{
  Same as Fig. \ref{fig:cool1} but for superfluidity class 4. \label{fig:cool4}}
\end{figure}

\begin{figure}
\resizebox{\hsize}{!}{\includegraphics{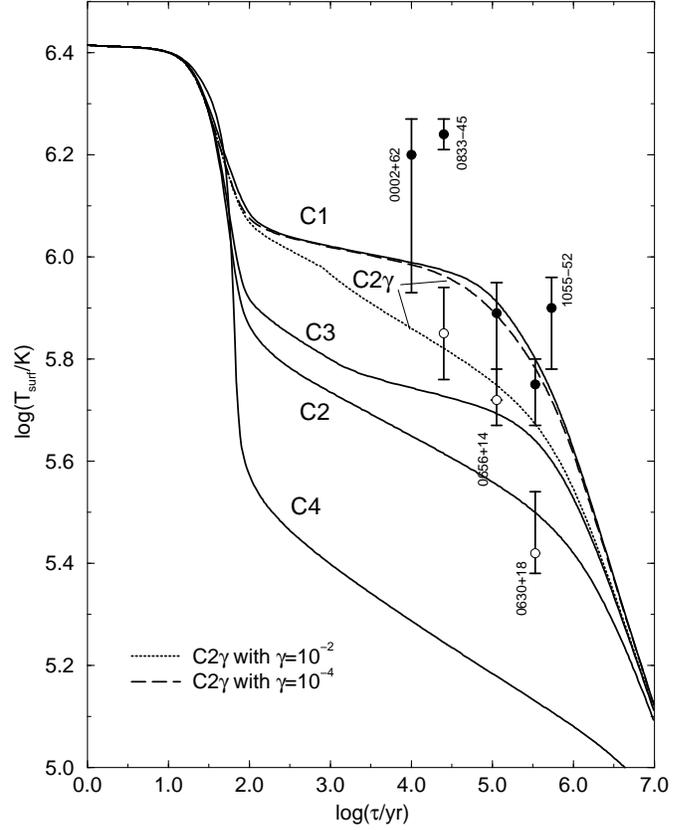}}
\caption[]{Thermal evolution of a $M=1.4M_\odot$ neutron star obtained 
for the HV$_{\rm pn}$ equation of state. The curves correspond to different 
superfluidity classes. \label{fig:cool5}}
\end{figure}

Among the soft X-ray observations of the 27 sources which were
identified as pulsars, the ROSAT and ASCA observations of PSRs
0002+62, 0833-45 (Vela), 0656+14, 0630+18 (Geminga) and 1055-52 (see
Table \ref{tab:observations}) achieved a sufficiently high photon flux
such that the effective surface temperatures of these pulsars could be
extracted by two- or three-component spectral fits
\cite[]{Oegelman95a}.  The obtained effective surface temperatures,
shown in Figs. \ref{fig:cool1} to \ref{fig:cool5}, depend crucially
on whether a hydrogen atmosphere is used or not. The kind of
atmosphere of individual pulsars could be determined by considering
multi-wavelength observations \cite[]{Pavlov96a}. Since the measured
spectra are however restricted to the X-ray energy band up to now, one
is not in the position to determine the kind of atmosphere. For that
reason we investigated both the blackbody model and the
hydrogen-atmosphere model, distinguished in Figs.  \ref{fig:cool1} to
\ref{fig:cool5} by error bars with a solid and hollow circle,
respectively.  All error bars represent the $2\sigma$ error range due
to the small photon fluxes.

\begin{table*}
\caption[]{Surface temperatures as measured by an observer at 
  infinity, $T_{\rm s}^\infty$, and spin-down ages, $\tau$, of
  observed pulsars. \label{tab:observations}}
\begin{tabular}{@{}ccccc}
\hline
Pulsar & $\log\tau$ [yrs] & Model atmosphere & $\log T_{\rm s}^\infty$ [K] 
& Reference \\
\hline
0002+62 & $\sim 4$\dag 		& blackbody             
& $6.20^{+0.07}_{-0.27}$        & \cite{Hailey95a}, 
				  Fig. 2 with lower limit on $N_{\rm H}$ \\
\hline
0833-45 & $4.4\pm 0.1$\dag	& blackbody             
& $6.24\pm 0.03$                & \cite{Oegelman95a}, Table III \\
(Vela)  &                  	& magnetic H-atmosphere 
& $5.85\pm 0.09$ 		& \cite{Page96a}, Fig. 1 \\
\hline
0656+14 & $5.05$ 		& blackbody
& $5.89^{+0.06}_{-0.17}$\ddag 	& \cite{Greiveldinger96a}, Table 2 \\
        &                  	& magnetic H-atmosphere 
& $5.72^{+0.06}_{-0.05}$\ddag	& \cite{Anderson93} \\
\hline
0630+18 & $5.53$ 		& blackbody
& $5.75^{+0.05}_{-0.08}$        & \cite{Halpern97a}, Table 2 \\
(Geminga)&                 	& magnetic H-atmosphere 
& $5.42^{+0.12}_{-0.04}$        & \cite{Meyer94},
				  Fig. 2a with $B_{12}=1.18$ \\
\hline
1055-52 & $5.73$ 		& blackbody
& $5.90^{+0.06}_{-0.12}$\ddag 	& \cite{Greiveldinger96a}, Table 2 \\
\hline
\end{tabular} 
\comments{\dag~~ estimated true age instead of spin-down age (see text). \\
	\ddag~~ The $2\sigma$ range is not given in the respective
	reference. We therefore estimated it by multiplying the given
	$1\sigma$ error with a factor 2.}
\end{table*}

Except for PSRs 0833-45 (Vela) and 0002+62, all ages are estimated by
their spin-down age $\tau=P/2\dot P$. This relation implies however
that both the moment of inertia and the magnetic surface field are
constant with time, and that the braking index $n$ is equal to its
canonical value 3 (angular momentum loss due to pure magnetic dipole
radiation). The true ages may therefore be quite different from the
spin-down ages. The age of Vela was recently determined by \cite{Lyne96a}, and the approximate age of PSR 0002+62 is given
by an estimate of the age of the related supernova remnant G 117.7+06
\cite[]{Hailey95a}.

All calculations were performed for a gravitational mass of
$M=1.4M_\odot$, about which the observed pulsar masses tend to
scatter. In Figs. \ref{fig:cool1} to \ref{fig:cool5}, we plot the
surface temperature of the neutron star models, as observed at
infinity, against the stars´ age. Figure \ref{fig:cool1} shows the
cooling bands for superfluidity case 1 and the equation of state
classes A--D (see Sect. \ref{sec:models} for the definition of the
classes). Superfluidity suppresses the nucleon processes so strongly,
that the bands of the three classes standard, intermediate and
enhanced nucleon cooling nearly coincide. Since the hyperon processes
are not suppressed within the models used here, the band of scenario
D1 is located much below the other three bands. Note that the
relatively low surface temperature, in comparison with other works
\cite[e.g.][]{Umeda94,Page95,Schaab95a}, of the standard scenario A1
is caused by the inclusion of the superfluid pair breaking and
formation process \cite[]{Voskresenskii87a,Schaab95b}, which is not
included in older works. The surface
temperature obtained by fitting with a magnetic hydrogen atmosphere
(hollow circles) lie right between the scenario D1 and the scenarios
A1--C1. In this model of superfluidity the gap between the bands can
only be filled by reducing the superfluid critical temperature of the
neutrons \cite[]{Page95,Schaab95a}.

By increasing the inner magnetic field strength above its critical
value we reach the second case of superfluidity
(Fig. \ref{fig:cool2}). Band C2, which corresponds to enhanced nucleon
cooling, moves to lower temperatures. The observation of PSR
0630+18 (Geminga) is now consistent with this scenario as long as this
pulsars has a magnetised hydrogen atmosphere.

If the magnetic field is in the order of the critical field strength
but still below it, a $m=2$--superfluid is first formed by the
neutrons, and later (as the star cools below $T_{\rm c,2}$, see
Eq. \ref{eq:crit.temp2}) this superfluid turns into a nodeless
state. This possibility is referred to by class 3
(Fig. \ref{fig:cool3}). Only in scenarios C3 and D3 the temperature
reaches this second critical value $T_{\rm c,2}$ during the neutrino
cooling era, that is during the time where cooling is dominated by
energy loss due to neutrino emission. For that reason the band of the
C3-scenario is pushed up again to higher temperature which are
consistent with the surface temperature obtained by fitting the
observations with a magnetised hydrogen atmosphere (apart from PSR
0630+18). Scenarios B3 and A3 show again the same cooling behaviour.

In case 4, we assume that the neutron superfluid does not extend
throughout the whole neutron star's core or the critical temperature
is very small $T_{\rm c}\lesssim 10^8$~K (Fig. \ref{fig:cool4}). Then
the nucleon processes are not suppressed everywhere, and particularly
the models of the C4-scenario of enhanced nucleon cooling cool down
very fast. Certainly this scenario, as well as scenarios D1--D4, is
not consistent with the observations, no matter what kind of
atmosphere is used. Case 4 would therefore clearly favour the standard
or the intermediate cooling scenarios. The higher emissivity of the
modified Urca process in scenario B4 lead now to considerably smaller
surface temperatures in comparison with scenario A4.

Finally we consider the case 2$\gamma$, where the critical field
strength $H_{\rm c}(T=0)$ is exceeded only inside the flux tubes. The
suppression factor $\mathcal R$ in Eq. (\ref{eq:gamma.suppr}) depends
on the volume fraction $\gamma$ occupied by the flux tubes. We study
the cooling behaviour for two values of $\gamma$: $\gamma =10^{-4}$,
which corresponds to a bulk magnetic field strength $H=2\times
10^{13}$~G (see Eq. \ref{eq:gamma.factor}), and $\gamma=10 ^{-2}$,
which corresponds to $H=2\times 10^{15}$~G. To clarify the impact of
this case on the cooling history of neutron stars, we plot cooling
tracks for the HV$_{\rm pn}$ equation of state instead of bands
(Fig. \ref{fig:cool5}). Since the neutrino emission is exponentially
suppressed only outside the proton flux tubes the effect of suppression
is not as high as in the case of model C1 (which corresponds to $\gamma
=0$), but still larger than in the model C2 (which corresponds to
$\gamma =1$). The obtained surface temperatures are consistent with
the surface temperature obtained by fitting the observations with a
magnetic hydrogen atmosphere (again apart from PSR 0630+18). Although
the surface temperature obtained in both scenarios C3 and C2$\gamma$ with
$\gamma=10^{-2}$ are consistent with the observation of PSR 0656+14
within the magnetic hydrogen atmosphere model, the bulk magnetic
field strength in both models differ by two orders of magnitude.

\section{Conclusions and discussion} \label{sec:disc}

In this work we have incorporated the possibility of nodes of the
neutron gap energy along the magnetic axis in the core of a neutron
star in simulations of thermal evolution. The realization of such
superfluids depends on a rather strong magnetic field, that might
however occur in the interior of a neutron star or in the interior of
flux tubes of the proton type II superconductor. In any case the state
of $m=2$--superfluid turned out to have several interesting
consequences.

Since the suppression factor $\mathcal{R}$ of neutrino emissivity
decreases only polynomially as function of $T/T_{\rm c}$ instead of
exponentially, the effect of superfluidity is not as large as in the
case of nodeless ($m=0$) neutron superfluidity, which was studied in
literature so far. Particularly the differences in the undiminished
neutrino emissivities of the cooling classes A--D remain visible in
the resulting surface temperatures.

The second important consequence is that in scenario C3 the second critical
temperature $T_{\rm c,2}$ depends on the internal magnetic field
(see. Eq. \ref{eq:crit.temp2}). In scenario C2$\gamma$ the suppression
factor depends on the factor $\gamma$ (see Eq. \ref{eq:gamma.suppr}),
which itself depends on the bulk magnetic field strength (see
Eq. \ref{eq:gamma.factor}). By varying the magnetic field the band
corresponding to scenarios C2$\gamma$ and C3 can be moved to lower or
higher surface temperatures in the age-range $10^2\,{\rm
yr}\lesssim\tau\lesssim 10^5\,{\rm yr}$. This can also be achieved by
varying the maximum energy gap, as it was done by \cite{Page95}
and \cite{Schaab95a}. Provided, that the energy gap
reaches its maximum value for densities smaller than the central
density of a typical neutron star, the maximum energy gap should be
the same for all neutron stars. This is different with respect to the
inner magnetic field strength since at least the surface magnetic
field strengths of individual pulsars differ by five orders of
magnitudes. This consequence may become important if two neutron stars
are found which have about the same age but different surface
temperatures, thought this might also be explained by different star
masses in connection with the intermediate cooling scenario \cite[see
Fig. 8 in][]{Schaab95b} and by more or less
accreted matter on the surface of the star \cite[see Fig. 9 in][]{Potekhin96c}. The two scenarios C2$\gamma$ and C3 differ
in the internal magnetic field strength which is needed to explain,
e.g., the surface temperature of PSR 0656+14 obtained within the
hydrogen atmosphere model. Whereas for scenario C3 the rather strong
magnetic field $H\sim 10^{17}$~G is needed, a bulk magnetic field
strength $H \sim 10^{15}$~G is sufficient in scenario C2$\gamma$.

The possibility of the direct nucleon Urca process becomes more and
more probable. To the best of our knowledge all
relativistic Hartree-, Hartree-Fock-, and
Br\"uckner-Hartree-Fock-equations of state have sufficiently large
proton fractions to allow for the direct Urca process. In the past
there were some doubt about the reliability of such high proton
fractions, because the nonrelativistic variational equations of state,
like \UVU\, and \UVT\, \cite[]{Wiringa88}, predict proton fractions
which are below the threshold value for the direct Urca process, since
the symmetry energy does not monotonously increase with density. This
seems to depend however on the used potentials, since more modern
potentials, like AV$_{18}$, Nijm-I,II, Reid93 and CD-Bonn, yield
higher proton fractions \cite[]{Engvik97a}.

It seems therefore reasonable to restrict our following discussion to
classes C and D. Since the impact of hyperon pairing has not been
investigated yet, let us assume that all observed neutron stars belong
to class C. Under this assumption the situation would become rather
enlightened if the chemical composition of the neutron stars envelope
could be determined by multi-wavelength observations. In the case that
the blackbody fits will turn out to be adequate only scenario C1 is in
agreement with the observations. If the magnetic hydrogen fits are
more reasonable only scenarios C2, C2$\gamma$, and C3 are consistent
with the data. However also scenarios C1 and C4 can be brought into
agreement if one reduces the gap energy in the case of C1
\cite[]{Page95,Schaab95a} or takes effects of an accreted envelope into
account in the case of C4 \cite[]{Potekhin96c,Page96b}. As it was
already claimed, these different scenarios could be further
distinguished, if the surface temperature of two or more neutron stars with similar ages could be determined.

\begin{acknowledgements}

We like to thank the referee for his valuable comments. One of us
(Ch.~S.) gratefully acknowledges the Bavarian State for
financial support. Tables with detailed references to the used
ingredients and with the obtained cooling tracks can be found on the
Web: http://www.physik.uni-muenchen.de/sektion/suessmann/astro/cool/schaab.0797.

\end{acknowledgements}



\end{document}